\documentclass[%
 reprint,
 superscriptaddress, 
 amsmath,amssymb,
 prl,                
]{revtex4-2}

\usepackage{graphicx}
\usepackage{dcolumn}
\usepackage{bm}

\begin{document}

\title{Harnessing electrostatics through temperature modulations to control ferroelectrics}

\author{Cameron Scott}
 \affiliation{Smart Materials Unit, Luxembourg Institute of Science and Technology (LIST), Esch-sur-Alzette, Luxembourg}

\author{Xabier Diaz de Cerio}%
\affiliation{Smart Materials Unit, Luxembourg Institute of Science and Technology (LIST), Esch-sur-Alzette, Luxembourg}

\author{Jorge {\'I}{\~n}iguez-Gonz{\'a}lez}%
\affiliation{Smart Materials Unit, Luxembourg Institute of Science and Technology (LIST), Esch-sur-Alzette, Luxembourg}
\affiliation{Department of Physics and Materials Science, University of Luxembourg, Belvaux, Luxembourg}

\begin{abstract}
Temperature modulations provide an alternative method for dynamically controlling the ferroelectric state. In this paper, we use scale-independent Landau potentials and predictive atomistic simulations to explore how temperature modulation can harness the depolarizing field to obtain non-electrical poling. We further predict that temperature gradients can be combined with strain to induce persistent polar textures such as multidomain states.

\end{abstract}

\maketitle

\begin{figure}
    \includegraphics[width=0.8\columnwidth]{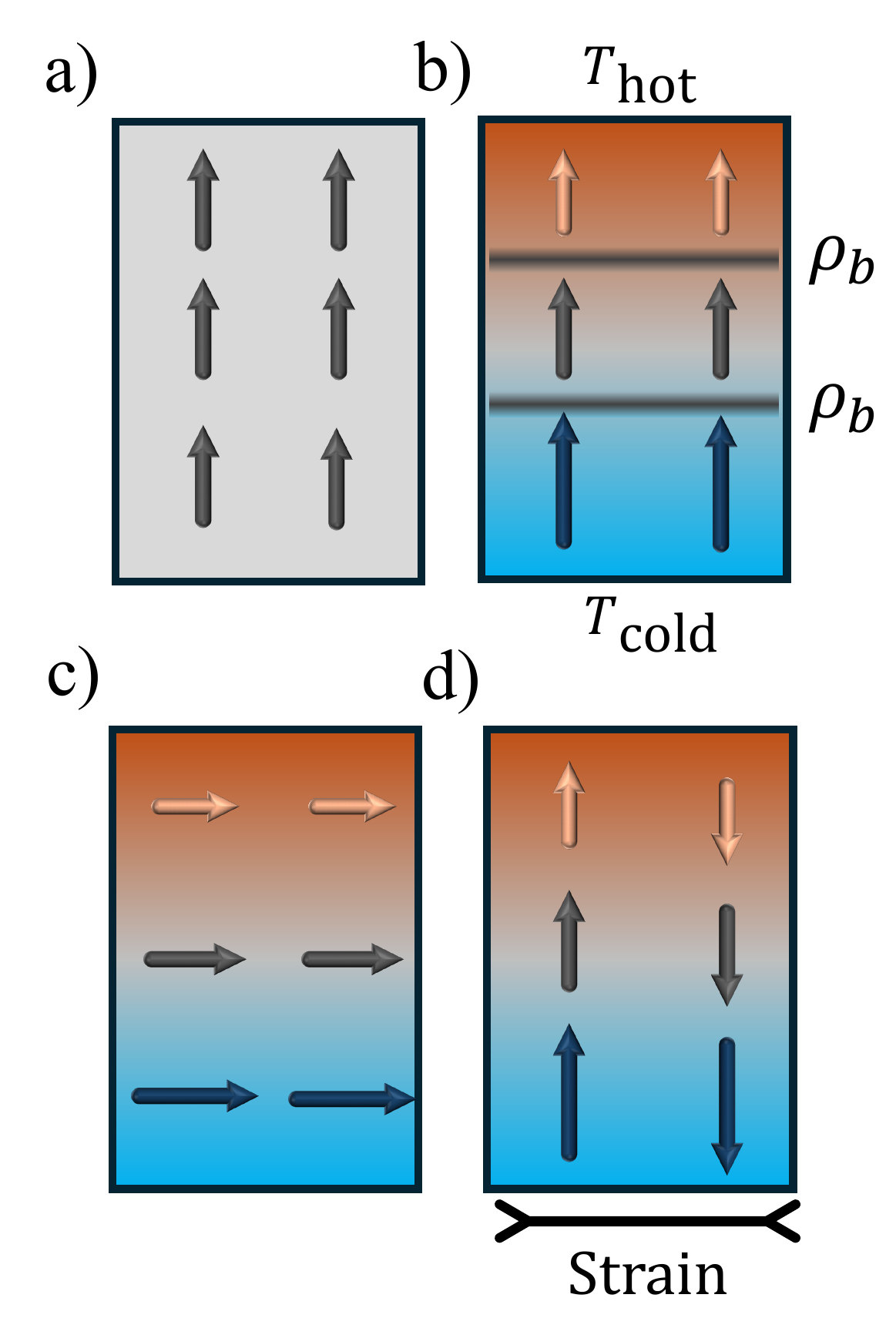}
    \caption{Rotation of dipoles under a temperature modulation. a) The ferroelectric monodomain. b) $T$-modulations and the pyroelectric effect leads to destabilising bound charges. c) Dipole rotation eliminates these bound charges. d) The inclusion of compressive strain promotes domain formation as another route to suppress the bound charge.}
    \label{fig:intro_figure}
\end{figure}

Dynamically controlling the state of ferroelectric materials is a challenge of fundamental and technological importance. The direct coupling of the polarization to an external electric field provides the most straightforward method for dynamical control, and even strain may offer a means for agile manipulation of freestanding ferroelectric membranes~\cite{chiabrera2022freestanding,hong2020extreme}. Temperature variations provide us with another obvious strategy to tune a ferroelectric's behavior. While temperature is usually varied homogeneously, there is promise for control via spatial temperature modulations ($T$-modulations), which will be our focus here.

The pyroelectric nature of all ferroelectrics means that a temperature gradient across a sample leads to an associated gradient in the polarization. This effect was first explored in 1985~\cite{strukov1985study}, when $T$-gradients were applied to ferroelectric triglycine sulfate crystals leading to a more diffuse phase transition and broader peak in the dielectric susceptibility.  Similar effects have been obtained through compositional gradients~\cite{choudhury2011geometric}, and various devices that make use of the internal potential (and shifted hysteresis loops) created by a polarization gradient, termed \textit{graded ferroelectric devices} \cite{mantese1995ferroelectric,mantese2005graded}, have been proposed. Further, it has been shown that the $T$-gradient induced by a moving, high-intensity laser can write ferroelectric domains in LiNbO$_{3}$ through a thermoelectric effect~\cite{xu2022femtosecond}. 

The mentioned graded ferroelectric devices must deal with a basic electrostatic effect: whenever there is a gradient in the polarization, bound charges ($\rho_b = - \nabla\cdot\mathbf{P}$) and the associated depolarizing field appear, suppressing polarization along the direction of the gradient. In this paper, we expand on this notion to explore such depolarizing effects, induced here by $T$-modulations. Our central ideas are depicted in Fig.~\ref{fig:intro_figure}. In Fig.~\ref{fig:intro_figure}(a) we have a typical ferroelectric compound with a well defined polar axis along the vertical direction, with all microscopic dipoles of equal magnitude aligned. When a $T$-modulation along the vertical direction is applied to the material, a polarization gradient emerges parallel to it (Fig.~\ref{fig:intro_figure}(b)), creating bound charges and a depolarizing field that penalizes the polarization oriented parallel to the gradient. Presumably, this should lead to a rotation of dipoles to be perpendicular to the $T$-modulation, so that the depolarizing field is eliminated (Fig.~\ref{fig:intro_figure}(c)). If, on the other hand, we are dealing with a uniaxial ferroelectric, the dipole rotation is not a possibility and the material will presumably break into domains so the depolarizing field is reduced (Fig.~\ref{fig:intro_figure}(d)). In what follows we present a numerical proof-of-concept for such control strategies.

We begin by testing our main hypothesis using an approximate Landau model of a ferroelectric subject to a $T$-modulation. For simplicity and without loss of generality, let us assume we work with a perovskite oxide like PbTiO$_{3}$. Let us imagine a sample consisting of two identical parts essentially corresponding to the hot and cold regions of Fig.~\ref{fig:intro_figure}, each described by the following sixth-order Landau potential,
\begin{equation}
\label{eq:landau}
\begin{split}
    F_k = & ( A_{2} + A_{2}' T_k ) ( P_{k,\perp}^2 + P_{k,\parallel}^2 ) \\
    & + A_{4} ( P_{k,\perp}^2 + P_{k,\parallel}^2 )^2 \\
    & + A_{22} P_{k,\perp}^2 P_{k,\parallel}^2 \\
    & + A_{6} ( P_{k,\perp}^2 + P_{k,\parallel}^2 )^3,
\end{split}
\end{equation}
where the index $k$ distinguishes between the hot and cold regions, and $P_{k,\perp}$ and $P_{k,\parallel}$ denote the polarization components perpendicular and parallel to the $T$-modulation, respectively. These two components are identical by symmetry in the absence of a $T$-modulation, and so share the same Landau coefficients. We then couple the two parts with the requirement that $D_\parallel$, the component of the electric displacement $\mathbf{D}$ parallel to the $T$-modulation, is continuous across the interface \cite{jackson2012classical}. The corresponding electrostatic energy is
\begin{equation}
\label{eq:electrostatic}
    F_{\text{elec}} = \frac{x_{\text{hot}} x_{\text{cold}}}{2\epsilon_0} ( P_{\text{hot},\parallel} - P_{\text{cold},\parallel} )^2,
\end{equation}
where $x_{\text{hot}}$ and $x_{\text{cold}}$ are the volume fractions of the two parts of our sample. This term models the effect of the bound charge and depolarizing field. The total model is then $F=x_\mathrm{hot}F_{\text{hot}} + x_\mathrm{cold}F_{\text{cold}} + 2 F_{\text{elec}}$, so the two parts of our model system are coupled only through electrostatics. Here we assume periodic boundary conditions along the $T$-modulation direction, which are equivalent to positioning the sample between two electrodes with perfect screening.

We now study the behavior of this model for a representative ferroelectric, PbTiO$_{3}$, using the Landau potential $\tilde{F}_{3}$ in Ref.~\onlinecite{pulzone2025machine}. This potential is derived from atomistic second-principles simulations and reproduces the experimental behavior of PbTiO$_{3}$ very well \cite{wojdel2013first} except for a relatively low Curie temperature (510~K vs. the experimental value of 760~K) that is not critical for the present purposes.

\begin{figure}
    \includegraphics[width=0.9\columnwidth]{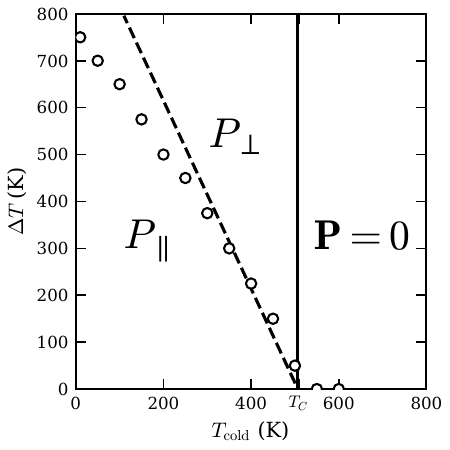}
    \caption{Phase diagram for temperature modulated ferroelectrics constructed using the Landau model of Eq.~\ref{eq:landau} demonstrating the rotation of dipoles with increasing temperature modulation. White dots denote the phase boundary as computed from second-principles molecular dynamics. We use $x_\mathrm{hot}=x_\mathrm{cold} = 0.5$. }
    \label{fig:phase_diagram}
\end{figure}

Our calculations lead to the phase diagram in Fig.~\ref{fig:phase_diagram}, presented as a function of $T_{\rm cold}$ and $\Delta T = T_{\rm hot} - T_{\rm cold}$. For $T_{\rm cold} > 510$~K our compound is in the paraelectric state. For $T_{\rm cold}< 510$~K, the global minimum of the free energy corresponds to the state with the polarization perpendicular to the temperature gradient (denoted $P_\perp$), which suffers from no electrostatic penalty. Finally, we find that the state with the polarization parallel to the temperature modulation ($P_\parallel$) is a local minimum of the free energy for sufficiently small values of $\Delta T$; the dashed line in Fig.~\ref{fig:phase_diagram} indicates its metastability limit.

Note that these results are consistent with the notion sketched in Figs.~\ref{fig:intro_figure}(a-c): If we start from a configuration with a vertical polarization, e.g. for $T_{\rm cold}=T_{\rm hot}= 300$~K, and then increase $T_{\rm hot}$, the polarization ($\P_\parallel$) is guaranteed to rotate toward the horizontal axis ($P_\perp$) as we approach $T_{\rm hot}\approx 700$~K (i.e., $\Delta T\approx 400$~K). Interestingly, in this example, the $P_\parallel$ state involves a hot region that remains polarized even though $T_{\rm hot}>T_{\rm C}$, on account of the electrostatic coupling with the cold part. Then, in the transformed state with $P_\perp$, we have ${\bm P}_{\rm hot} = 0$. It is also worth noting that the boundary between the  regions with $P_\perp$ and $P_\parallel$ slopes downwards as $T_\mathrm{cold}$ approaches the bulk transition temperature and, consequently, the polarization rotation can be driven with a progressively weaker modulation.

\begin{figure*}
    \includegraphics[width=0.9\textwidth]{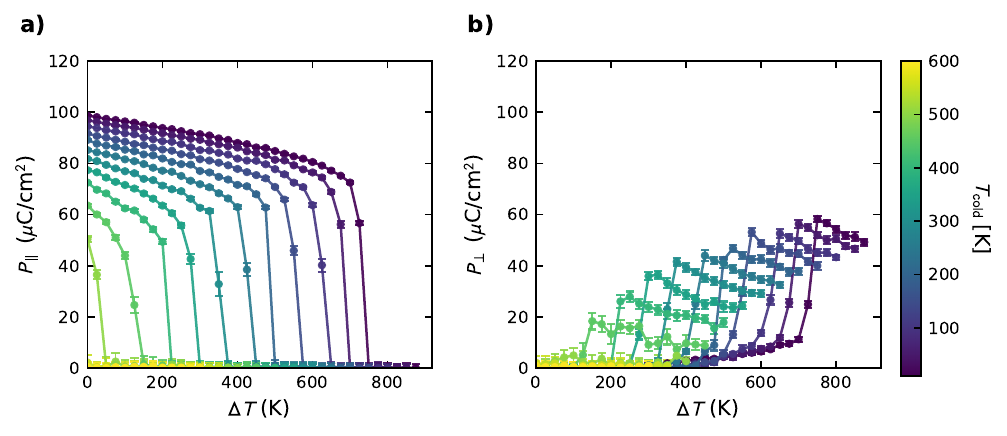}
    \caption{Polarization rotation under a sinusoidal temperature modulation using Langevin molecular dynamics with damping constant of $\gamma = 1000 fs^{-1}$. a) $P_\parallel$ is destroyed with sufficiently $T$-modulation and b) $P_\perp$ appears at the same value of $\Delta T$.}
    \label{fig:scup_rotations}
\end{figure*}

To test these conclusions, we use molecular dynamics (MD) simulations based on the atomistic second-principles potential from which our Landau model for PbTiO$_{3}$ was derived~\cite{pulzone2025machine,wojdel2013first}. Specifically, we run Langevin MD in supercells that are a $10\times10\times20$ repetition of the elemental 5-atom perovskite cell, and with the polar axis along the $z$ direction. To implement the temperature modulation, we adapt the Langevin equation of motion by assigning a temperature $T_i$ to the $i$th atom,
\begin{equation}
    m_i \frac{d^2 \mathbf{r}_i}{dt^2} = \nabla U(\mathbf{r}_i)- \gamma m_i \frac{d \mathbf{r}_i}{dt} + \sqrt{2 m_i \gamma k_B T_i} \boldsymbol{\eta}_i(t).
\end{equation}
We apply a sinusoidal modulation with extrema $T(z=\frac{L_z}{4}) = T_\mathrm{cold}$ and $T(z=\frac{3L_z}{4}) = T_\mathrm{hot}$, where $L_{z}$ is the vertical dimension of our simulation supercell. 

We begin by fixing $T_\mathrm{hot}=T_\mathrm{cold}$ and running MD for 100~ps with a 4~fs timestep. We ignore the first 50~ps to account for thermalization, and extract the averages of the polarizations and strains from the remaining 50~ps. Starting from the final structure of this MD run, we then increment $T_\mathrm{hot}$ (keeping $T_\mathrm{cold}$ fixed) and repeat the MD run. We checked that these calculation conditions yield results that are sufficiently converged for the present purposes.

\begin{figure}
    \includegraphics[width=0.9\columnwidth]{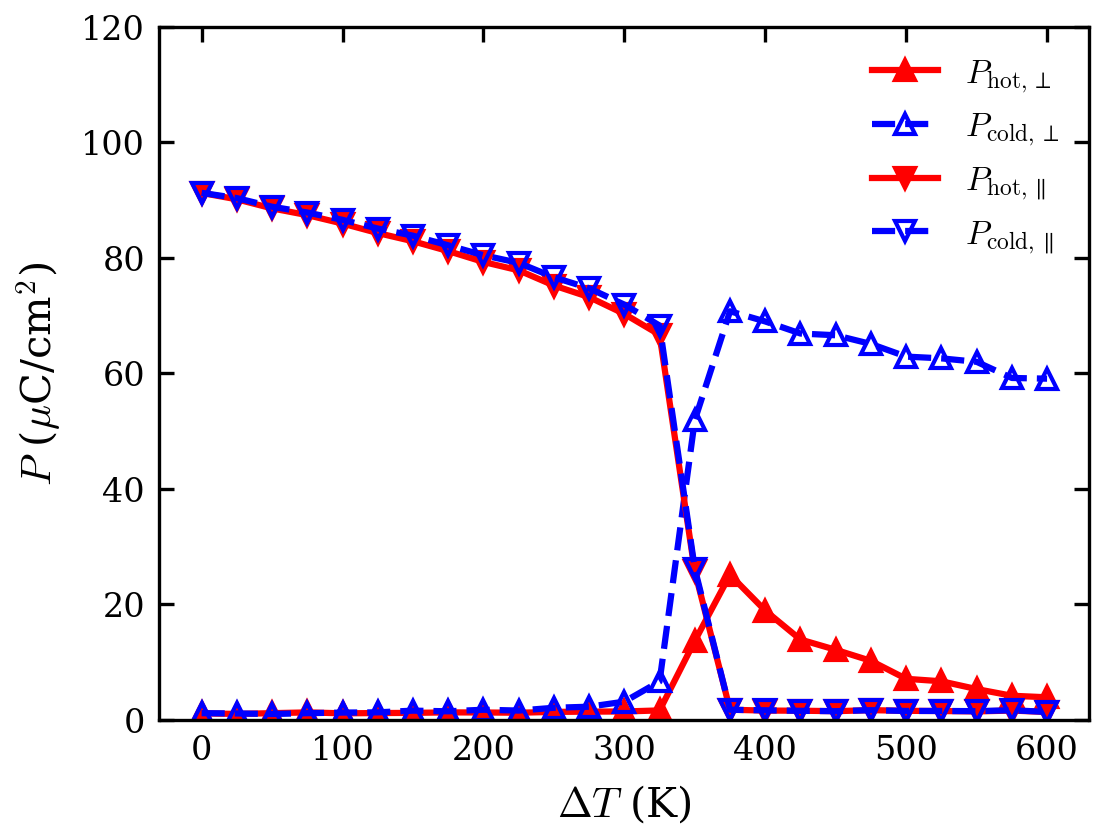}
    \caption{Rotation of dipoles under a sinusoidal temperature modulation with  $T_{\rm cold} = 300$~K. Polarisation calculated seperately for the hot and cold regions of the of the simulation supercell.}
    \label{fig:rotation}
\end{figure}

Figure~\ref{fig:scup_rotations}(a) shows how $P_\parallel$ evolves under increasingly strong $T$-modulations. For each $T_\mathrm{cold}$, a large enough modulation forces the rotation from $P_\parallel$ to $P_\perp$. Indeed, Fig.~\ref{fig:scup_rotations}(b) demonstrates that $P_\perp$ appears as soon as $P_\parallel$ vanishes, confirming the essential picture discussed above from our simple Landau model. This is emphasized for a specific choice of $T_{\rm cold}$ in Fig.~\ref{fig:rotation}.

\begin{figure*}
    \includegraphics[width=\textwidth]
    {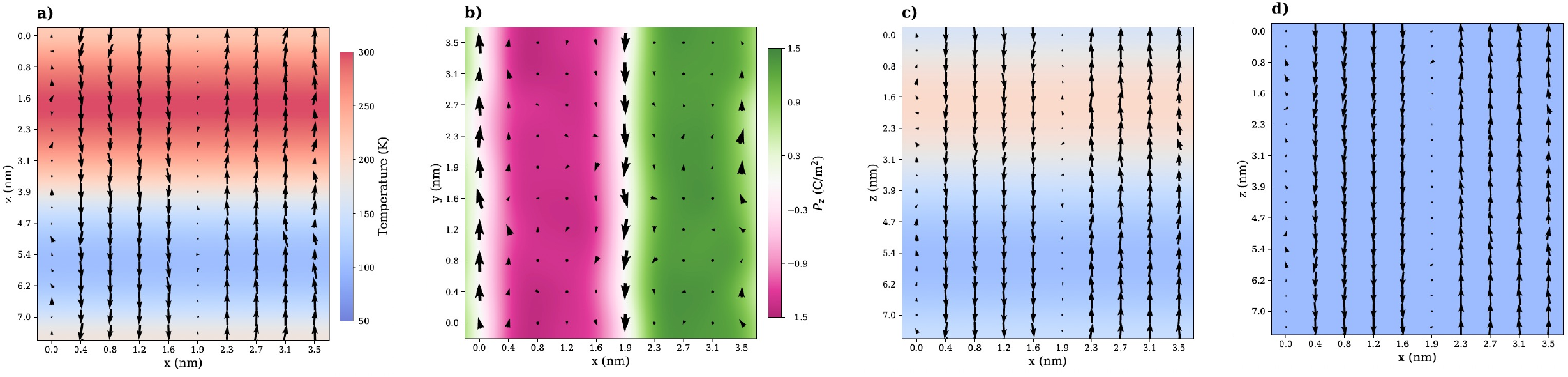}
    \caption{The combination of compressive strain and $T$-modulations lead to formation of persistent stripe domains in PbTiO$_3$. Panel $a$) shows an $xz$ slice of the supercell, with the color map  indicating the temperature modulation along $z$. Panels $b$) show a $xy$ slice (in the middle of cold region), with the color map indicating the $z$-component of polarization.  Panels $c)$ and $d)$ are similar to panel $a)$, and they demonstrate that the stripe domains persist even when the temperature modulation vanishes.}
    \label{fig:strain}
\end{figure*}

In fact, from Fig.~\ref{fig:scup_rotations} we can extract the phase boundary between the $P_\perp$ and $P_\parallel$ configurations, and overlay the results on the Landau phase diagram of Fig.~\ref{fig:phase_diagram}. Our nanoscale atomistic simulations are in near perfect agreement to the scale-independent Landau model, differing only at high temperatures where increased thermal fluctuations allow rotation before the vanishing of the $P_\parallel$ minimum. It is also interesting to note that our atomistic simulations confirm the possibility of having states with $P_\parallel\neq 0$ for $T_{\rm hot}>T_{\rm C}$, on account of the electrostatic coupling with the cold part.

Admittedly, the gradients considered in our atomistic simulations are very large to be of practical relevance, as they reach values of the order of 100~K/nm. But then, interestingly, the agreement between the atomistic results and the estimate based on a simple Landau model emphasizes that electrostatic effects like these can be expected to be scale free, and should apply for experimentally accessible $T$-modulations.

The above results suggest a different form of control of the ferroelectric state: if we cool down our system through $T_{\rm C}$ while subject to a $T$-modulation, we should be able to favor the polar axis perpendicular to said modulation. To test this, we reduce $T_\mathrm{cold}$ across $T_{\rm C}$ while applying a constant $\Delta T$ to establish the sinusoidal $T$-modulation, allowing the system to come to equilibrium before each successive temperature decrement. Once sufficiently below the transition temperature (at 475~K), we measure the orientation of the electric dipoles. To obtain meaningful statistics, we repeat this experiment 100 times for each $\Delta T$ considered. We find that, for $\Delta T=50$~K, the probability to observe the polarization perpendicular the $T$-modulation is 77~\%, significantly higher than 66~\% expected from random chance. This probability increases monotonically with increasing $\Delta T$, reaching 97~\% for $\Delta T=100$~K. Interestingly, we also find that $T$-modulations involving steeper gradients enhance this effect; in particular, a sharp step-like modulation yields a probability of 96~\% for $\Delta T = 50$~K. 

Now, let us consider the possibility sketched in Fig.~\ref{fig:intro_figure}(d), where the polarization is not allowed to rotate, which we hypothesize may lead to a multidomain configuration~\cite{bratkovsky2002formation}. This could be the case of uniaxial materials like LiNbO$_3$, and a similar situation pertains to ferroelectric perovskite thin films, e.g. of PbTiO$_{3}$, subject to an in-plane compressive strain~\cite{schlom2007strain}. In our atomistic simulations, we can mimic this situation by fixing the in-plane lattice vectors of the simulation supercell so that our simulated PbTiO$_{3}$ is forced to present a vertical polarization. A natural choice is to use the in-plane lattice constants of bulk PbTiO$_3$ in the ferroelectric phase (3.94~$\mathrm \AA$ according to our model), which corresponds approximately to the experimental case of growing a PbTiO$_3$ thin film on a (001) SrTiO$_3$ substrate. Figures~\ref{fig:strain}(a-b) show the results obtained under such an elastic constraint, subject to a sinusoidal temperature modulation with $T_{\rm cold}=100$~K and $T_{\rm hot}=300$~K: we observe the formation of electric stripe domains which persist as the temperature difference is reduced (Figures~\ref{fig:strain}(c-d)). In this way, $T$-modulation becomes a dynamical route towards creating persistent textures in ferroelectric materials.

Let us note that, for all these effects to occur, the temperature modulation must be present for at least the amount of time it takes for the dipole reorientation to be complete - about 20-50 ps in our simulations. Using the thermal conductivity of PbTiO$_3$ (approximately 4 ${\rm Wm^{-1}K^{-1}}$)\cite{martinez2004dielectric} and solving a one-dimensional heat equation, we estimate that a linear dimension greater than 50 ~nm is required for temperature differences of approximately 50~K to persist for longer than a few hundred picoseconds. Hence, thanks to the relatively low thermal conductivity of ferroelectric oxides, it seems viable to obtain the effects discussed here even in the nanoscale regime.

It is worth noting that our calculations of temperature induced dipole rotation may shed light on recent results on how spatial temperature modulations (with a $\Delta T=10$~K) impact the ferroelectric phase transition in BaTiO$_3$ \cite{tsukada2023temperature}. The authors of Ref. \cite{tsukada2023temperature} find that the sample is paraelectric close to the heat source. Then, interestingly, as the colder heat sink is approached, the structure transitions to a tetragonal phase with the polarization perpendicular to the $T$-modulation, which is clearly reminiscent of the phenomena described here. Then, the authors also find an alignment parallel to the $T$-modulation close to the heat sink. This probably reflects a thermoelectric effect \cite{xu2022femtosecond} that seems absent in our present simulations, potentially on account of our using periodic boundary conditions.

In conclusion, our results show how temperature modulation can be an effective method to control the ferroelectric state, allowing for the reorientation of electric dipoles, the selection of a particular polar axis or the creation of persistent polar textures in simple systems. Importantly, the results discussed here depend on long range electrostatic interactions and are largely independent of the size of the sample and details of the temperature modulation. In fact, they should apply as long as the temperature modulation can be sustained for long enough time to allow for the reconfiguration of the electric order. We thus propose temperature modulation as an alternative means for dynamical control of ferroelectricity and hope that these results will prompt experimental consideration of these ideas.

We extend our gratitude to Riccardo Hertel and Théo Bruant for helpful discussions. This work was supported by the Luxembourg National Research Fund (FNR, grant INTER/ANR/24/18960894/TOPOTHERM) and the French National Research Agency (ANR, grant ANR-24-CE30-5392), within the framework of the ANR–FNR 
collaboration TOPOTHERM.

The data used in this work is publicly available and can be found at 10.5281/zenodo.20826820.

\bibliography{apssamp}

\end{document}